# Double solution with chaos:
# Dynamic redundance and causal wave-particle duality

A.P. KIRILYUK*

Institute of Metal Physics, Kiev, Ukraine 03142

ABSTRACT. A system of two interacting, physically real, initially homogeneous fields is considered as the most elementary possible basis for the world construction in which one of them, a 'protofield' of electromagnetic nature, is attracted to another protofield, or medium, responsible for the eventually emerging gravitational effects. The interaction process is analysed with the help of the generalised 'effective (optical) potential method' in which we avoid any usual perturbative reduction. It then appears that for generic system parameters the protofields, instead of simply 'falling' one onto another and forming a fixed 'bound state', are engaged in a self-sustained process of nonlinear pulsation, or 'quantum beat', consisting in unceasing cycles of self-amplified auto-squeeze, or 'reduction', of a portion of the extended protofields to a very small volume followed by the inverse phase of extension. The centres of consecutive reductions form the physical 'points' of thus emerging, intrinsically discrete space, and each of them is 'selected' by the system in a causally random fashion among many equally possible versions (or 'realisations'). This is a manifestation of the universal phenomenon of dynamic redundance forming the unified basis of dynamically complex (chaotic) behaviour of any real system with interaction (physics/9806002). In both particular and general cases, this universally applicable and causally specified picture of dynamic complexity is obtained not as a result of some artificially imposed 'principle' or 'influence', but only due to the unreduced, universally nonperturbative analysis of a generic interaction process. The spatially chaotic sequence of the protofield reduction events constitutes the elementary physical 'clock' of the world providing it with the causally emerging, unceasing, and intrinsically irreversible time flow. The complex-dynamical quantum beat process in the system of interacting protofields is observed as the elementary particle (exemplified by the electron) possessing the causally consistent, physically real property of wave-particle duality that actually completes the double solution concept of Louis de Broglie with the unreduced dynamical chaos (quant-ph/9911107). The latter can be considered as the direct extension of de Broglie's idea about the 'hidden thermodynamics of the isolated particle'. These rigorously derived results form the basis for the causally complete, intrinsically nonlinear version of quantum (wave) mechanics which gives physically and mathematically consistent, dynamically based explanations for such fundamental 'quantum' properties, 'inexplicable' within the existing interpretations, as (now) physically real wavefunction (including de Broglie's 'matter wave'), 'corpuscular' (localised) state of the elementary fields/particles, wave-particle duality, quantum indeterminacy, and discreteness (quantization) (see also quant-ph/9902016).

―――――――――――――
*Address for correspondence: Post Box 115, Kiev-30, Ukraine 01030.
 E-mail: kiril@metfiz.freenet.kiev.ua

RÉSUMÉ. *Le système de deux champs physiquement réels, interagissant entre eux et initialement homogènes est étudié comme la base la plus élémentaire possible de la structure du monde, où un des champs, le 'protochamp' de nature électromagnétique, est attiré vers l'autre protochamp, ou 'milieu', responsable pour les effets de gravitation éventuellement émergent. Le processus de l'interaction est analysé à l'aide d'une généralisation de la 'méthode du potentiel effectif (optique)' dans laquelle nous évitons les réductions perturbatives habituelles. Il s'ensuit alors, pour les valeurs génériques de paramètres du système, que les protochamps, au lieu de 'tomber' l'une sur l'autre et former un 'état lié' fixe, s'engagent dans le processus dynamiquement maintenu de pulsation nonlinéaire, ou 'battement quantique', qui consiste en cycles répétés de l'auto-contraction, ou 'réduction', d'une partie des protochamps étendus jusqu'à un très petit volume suivie par la phase inverse de leur extension. Les centres des réductions successives forment les 'points' physiques de l'espace ainsi émergent, intrinsèquement discret, et chaque centre est 'sélectionné' par le système entre ses plusieurs versions (ou 'réalisations') également possibles dans la manière qui est donc causale et aléatoire (probabiliste) à la fois. On a affaire ici avec manifestation du phénomène universel de la redondance dynamique qui accompagne chaque processus d'interaction générique et constitue la base du comportement dynamiquement complexe (chaotique) qui en résulte (physics/9806002). Ce tableau causal, universellement applicable et bien spécifié de la complexité dynamique est obtenu, dans le cas particulier aussi que général, non pas comme résultat d'un 'principe' ou 'influence' imposé artificiellement, mais uniquement grâce à l'analyse nonréduite, universellement nonperturbative du processus d'interaction générique. La séquence d'événements de réduction des protochamps constitue 'l'horloge' physique élémentaire du monde et lui donne le cours de temps réellement émergent, incessant, et intrinsèquement irréversible. Le processus dynamiquement complexe de battement quantique maintenu dans le système des protochamps par leur interaction est observé comme une particule élémentaire (présenté par l'électron) qui possède naturellement la propriété de dualité onde-corpuscule physiquement réelle, cohérente et effectivement complétant le concept de la double solution de Louis de Broglie par le chaos dynamique non-réduit (quant-ph/9911107). Ce dernier peut être considéré comme l'extension directe de l'idée Broglienne de la 'thermodynamique cachée de la particule isolée'. Ces résultats rigoureusement déduits constituent la base pour la version complète et causale de la mécanique quantique (ondulatoire) intrinsèquement nonlinéaire qui propose des explications physiquement et mathématiquement cohérentes, dynamiquement basées pour les propriétés 'quantiques' fondamentalement 'inexplicables' dans les interprétations existantes, tels que la 'fonction d'onde' physiquement réelle maintenant (y compris 'l'onde de la matière' de Louis de Broglie), l'état 'corpusculaire' (localisé) des champs/particules élémentaires, dualité onde-corpuscule, incertitude quantique et discontinuité (quantification) (voir aussi quant-ph/9902016).*

# 1. Introduction: Incompleteness of causal wave mechanics and the involvement of (physical) dynamic complexity

The physical incompleteness of the 'standard', or Copenhagen, interpretation of quantum mechanics is a well-known fact, actually explicitly acknowledged by that paradigm itself and even forming one of its basic points [1-4]. The resulting fundamentally 'inexplicable' character of the most important properties of quantum object behaviour evokes an ever growing number of attempts to 'ameliorate' the fundamentals of quantum theory up to a reasonably 'complete', self-sufficient understanding containing no major contradictions (see e. g. [2-8]). An *a priori* most consistent approach underlies a group of *causal*, or *realistic* interpretations of quantum mechanics which try to describe the actual physical reality behind the standard quantum postulates 'seriously' [3], i. e. in principle in the same fashion as in various domains of macroscopic, 'classical' physics, where a well-specified qualitative description in terms of the participating 'tangible' entities (fields, particles, structures) should precede and further accompany the necessary mathematical presentation. A really complete, 'first-principle' causal theory should even *deduce* all the known properties of quantum behaviour, summarised in the standard postulates, as universal, physically transparent *emergent* features of dynamics of a small number of 'material' elementary entities like uniform fields. Such are the ambitions of the essentially nonlinear *wave mechanics* (*la mécanique ondulatoire*) initiated by Louis de Broglie from the very birth of quantum mechanics as one of its major starting points (de Broglian 'matter waves') [9,10], but also further developed by him and his collaborators as the (eventually) *causally complete contents* of the postulated external *scheme* of the standard interpretation [11-18]. Taking this way, one tries to obtain *not* an *interpretation* of a theory, but the most complete, *unreduced* version of the *theory itself*, equivalent to the *full, absolutely causal understanding* of the fundamental levels of reality, without any specific 'quantum mysteries', imposed formal 'postulates', or internal contradictions.[1] Note that such renowned founders of modern physics as H.

---

[1] In this sense one should make a clear distinction between the unreduced causality of de Broglie's approach and the explicitly simplified, auxiliary 'pilot-wave interpretation' proposed initially by de Broglie [19] as a concession to the emerging standard interpretation (see e. g. [22]), but later rediscovered by David Bohm [20], and now often presented as the main version of causal quantum mechanics under the name of 'de Broglie-Bohm approach/interpretation' [4,8,21]. Although the pilot-wave description and its various versions generally remain in the same causal 'camp' of *attempts*, they all definitely belong to the realm of *interpretations*, i. e. basically only external *reformulations*, of the *same* standard theory and level of understanding, with all its fundamental deficiencies; an interpretation does not introduce a major *qualitative*, physically meaningful extension which can alone provide a transition to the really complete understanding.

Lorentz, M. Planck, E. Schrödinger, and A. Einstein were also among staunch supporters of the unreduced causality in quantum theory [22].

However, the program of Louis de Broglie has remained incomplete, even though the particular results obtained, starting already from the *unreduced* justification [9] of the 'standard' expression for the 'de Broglie wavelength', now totally excluded from the majority of quantum theory presentations, definitely show that the hope to obtain the completely causal description is not vain, since there 'should be something' more specified and revealing a universal truth behind the found relations among the most fundamental principles of physics [10-17]. The achievements of de Broglie's theory are centred around the idea of the so called 'double solution' implying that an elementary particle should be described by *coexistence* of two qualitatively different states of a real physical field, a highly concentrated 'hump' with *corpuscular* properties and a smooth, *undulatory* distribution of the hump position related to low-magnitude 'tail' around the hump and corresponding to the canonical 'wavefunction' [10-13], the whole construction being simply a causally specified expression of the undeniable wave-particle duality. The related paradigm of the 'hidden thermodynamics of the isolated particle' [14-16] develops the double solution idea in a natural way by suggesting that, in agreement with the observed 'nonlocality' of quantum particle behaviour, the hump should permanently perform irregular, 'thermal' wandering within the accessible wavefunction domain. Despite its intuitively appealing, physically transparent structure and the related indeed promising fundamental relations, the double solution was never explicitly obtained, while the origin of the 'hidden thermodynamics' remains rather obscure, and everything shows that these difficulties have the same fundamental origin as wave-particle duality and quantum indeterminacy in general.

In this paper we describe a new, independent and really first-principle analysis of the elementary object dynamics which leads, however, to the results strikingly resembling the predicted properties of the double solution [10-13]. Therefore we can state that we have *explicitly* obtained the unreduced, causally completed version of the double solution realising the universal, essentially nonlinear, self-consistent, and realistic description of the most fundamental levels of being that was initiated by Louis de Broglie 75 years ago and then persistently, and considerably, developed despite the enormous objective and subjective difficulties [17, 18, 22]. The consistency of the obtained solution is confirmed by the fact that it emerges in natural unity with the causally extended, 'physically' explained (special and general) relativity and the related causal versions of the notions of space, time, mass-energy, electric charge and spin [25]. This decisive progress would be impossible without some qualitative novelty in the method of analysis which takes the form of the *dynamic redundance (or*

*multivaluedness) paradigm* [23-25]. This basic extension of the notion of existence itself is rigorously *deduced* using a *universally* nonperturbative analysis of the 'method of effective dynamical functions' (a generalisation [26] of the known 'optical potential method' [27]). It is shown that *any* nontrivial interaction process leads to emergence of *many* complete and therefore *incompatible* compound system states, or *realisations* which should permanently, 'spontaneously' and *chaotically* (unpredictably) *replace* each other, and this provides the natural (and universal) basis for the *causal (dynamic) randomness* (or uncertainty) that can be described in terms of the *universal dynamic complexity* [23-25]. It is clear that the *unceasing, completely autonomous, chaotic* realisation *change* explicitly *obtained* as a universal *dynamic* phenomenon opens quite new possibilities for causal understanding of the 'mysterious' *coexistence* of the opposite types of states within the same 'quantum-mechanical' micro-object and the corresponding double solution. Below we present the application of the universal nonperturbative analysis to description of the elementary field behaviour, where the wave-particle duality and 'quantum' indeterminacy emerge as purely dynamical, causally inevitable properties within the obtained 'double solution with chaos'.

## 2. Causal randomness in a system with interaction, quantum beat, and the emergent wave-particle duality

Consider a system of two interacting, initially practically homogeneous (structureless), but physically real 'protofields' (or primal 'media') as a minimum possible basis of the world construction, and let us analyse the unreduced result of the interaction process development by avoiding any 'perturbative' cutting of the emerging events and structures. It is natural to suppose that one of the fields is of electromagnetic (e/m) origin/type (so that certain its perturbations are observed as the omnipresent e/m waves or photons), while the other one forms a 'gravitational background/medium' eventually responsible for gravitation (we shall see that it cannot be *directly* perceivable, in any its state). Being physically real, each (proto)field possesses an internal structure which determines its finite compressibility and can transmit perturbations, but whose elements are not resolvable at any level *within* our world due to the fundamental discreteness of any real system dynamics and the ensuing finite separation between the emerging 'levels of complexity' (we shall self-consistently justify this assumption later). In other words, the unperturbed protofields form a kind of 'ether' which is both physically real and unobservable because of its *actual* absolute homogeneity in terms of the well separated, 'coarse-grained' dynamics of the observable world, whereas *perturbations* of such ether *can* be observed and actually form elementary 'measured objects' and 'measuring instruments'. Both protofields are present 'everywhere' (the observed three-dimensional space dynamically *emerges* from their interaction and will be specified in further description), and we suppose that the e/m protofield is *attracted* to the gravitational background (one can consider also that it is attracted to, and tries to 'fall' over, a sufficiently rigid, omnipresent 'border of the world' playing thus the role of the gravitational protofield, probably together with a 'world at the other side'); indeed, only attractive (binding) interaction can provide a self-sufficient (closed) world dynamics (we can also call this basic interaction 'electro-gravitational coupling').

The compound system of interacting fields can be described by the *existence equation* of a quite general form which simply reflects the fact that the interacting entities (protofields) develop into a *single, integrated* (holistic) *system* characterised by a single *state-function* $\Psi(q,\xi)$ and generalised *eigenvalue E*:

$$[h_e(q) + h_g(\xi) + V_{eg}(q,\xi)]\Psi(q,\xi) = E\Psi(q,\xi) , \qquad (1)$$

where $q$ is the electromagnetic and $\xi$ the gravitational degrees of freedom, $h_e(q)$ and $h_g(\xi)$ are the generalised Hamiltonians (or any other actually measured functions) for the 'free' (non-interacting) media, $V_{eg}(q,\xi)$ is a generally unspecified (though eventually attractive and binding) interaction potential between the fields of $q$ and $\xi$, and $E$ is the energy (or other property corresponding to the selected 'generalised Hamiltonian' and representing a measure of dynamic complexity fixed at its lower level, see further analysis). We emphasize that the existence equation, eq. (1), has a very general, almost symbolical sense stating only that 'the system exists as a whole', which corresponds to the simplest possible construction of the described physical reality consisting of two interacting entities. This means also that this equation and the results of its analysis are applicable to *any* system of *arbitrary* interacting entities/objects, and such uniformity is important for understanding of the self-consistent, quasi-autonomous further development of the holistic world structure in the form of the 'universal hierarchy (arborescence) of dynamic complexity' [25] which does not contain any 'bizarre' rupture between 'quantum' and 'classical' levels of being (we specify this statement later).

For further analysis it would be useful to present the existence equation in the form of a system of equations by first expanding the state-function to be found, $\Psi(q,\xi)$, over a complete system of eigenfunctions, $\{\phi_n(q)\}$, of the free e/m field Hamiltonian, $h_e(q)$:

$$h_e \phi_n(q) = \varepsilon_n \phi_n(q) , \qquad (2)$$

$$\Psi(q,\xi) = \sum_n \psi_n(\xi)\phi_n(q) . \qquad (3)$$

The chosen complete system of functions corresponds to the directly observed, or *measured* (in an extended sense), entity and its main property which are supposed to be, in our case, an eventually obtained localised state of the e/m medium and its *space* position (a *physical* 'space point', see below), so that $\phi_n(q)$ can be close to $\delta$-functions, $\phi_n(q) \propto \delta(q - q_n)$, $\varepsilon_n$ is reduced to the point coordinate, $\varepsilon_n = q_n$, and eq. (2) reproduces the main property of the $\delta$-function [24,25]. However, this choice can be varied and we shall continue our analysis in its most general form incorporating such particular details which will be evoked where necessary.

After substitution of eq. (3) into eq. (1) and standard technical manipulations we obtain the equivalent system of equations in which the 'measured' e/m field variable, $q$, is replaced by the equivalent numbers, $n$, of the respective free-field eigen-modes (or 'points'):

$$[h_g(\xi) + V_{00}(\xi)]\psi_0(\xi) + \sum_n V_{0n}(\xi)\psi_n(\xi) = \eta\psi_0(\xi) , \qquad (4a)$$

$$[h_g(\xi) + V_{nn}(\xi)]\psi_n(\xi) + \sum_{n'\neq n} V_{nn'}(\xi)\psi_{n'}(\xi) = \eta_n\psi_n(\xi) - V_{n0}(\xi)\psi_0(\xi) , \qquad (4b)$$

where

$$\eta_n \equiv E - \varepsilon_n , \qquad (5)$$

$$V_{nn'}(\xi) \equiv \int_{\Omega_q} dq\, \phi_n^*(q) V_{eg}(q,\xi) \phi_{n'}(q) . \qquad (6)$$

and we have separated the equation with $n = 0$, while assuming that $n \neq 0$ and $\eta \equiv \eta_0$. Now we try to 'solve' the system of equations (4) by expressing $\psi_n(\xi)$ through $\psi_0(\xi)$ from eqs. (4b) with the help of the standard Green function technique [23-27] and inserting the result into eq. (4a), after which we obtain the single effective existence equation for $\psi_0(\xi)$ explicitly expressed only through the gravitational degrees of freedom ($\xi$) that play the role of a (generalised) *instrument of measurement*:

$$[h_g(\xi) + V_{\text{eff}}(\xi;\eta)]\psi_0(\xi) = \eta\psi_0(\xi) , \qquad (7)$$

where the *effective (interaction) potential* (EP), $V_{\text{eff}}(\xi;\eta)$, is given by

$$V_{\text{eff}}(\xi;\eta) = V_{00}(\xi) + \hat{V}(\xi;\eta) , \quad \hat{V}(\xi;\eta)\psi_0(\xi) = \int_{\Omega_\xi} d\xi' V(\xi,\xi';\eta)\psi_0(\xi') , \qquad (8a)$$

$$V(\xi,\xi';\eta) \equiv \sum_{n,i} \frac{V_{0n}(\xi)\psi_{ni}^0(\xi) V_{n0}(\xi')\psi_{ni}^{0*}(\xi')}{\eta - \eta_{ni}^0 - \varepsilon_{n0}} , \quad \varepsilon_{n0} \equiv \varepsilon_n - \varepsilon_0 , \qquad (8b)$$

and $\{\psi_{ni}^0(\xi)\}$, $\{\eta_{ni}^0\}$ are the complete sets of eigenfunctions and eigenvalues for an 'auxiliary', truncated system of equations (where $n \neq 0$)

$$[h_g(\xi) + V_{nn}(\xi)]\psi_n(\xi) + \sum_{n' \neq n} V_{nn'}(\xi)\psi_{n'}(\xi) = \eta_n \psi_n(\xi) \ . \tag{9}$$

The complete solution of the initial existence equation, eq. (1), is then obtained in the form

$$\Psi(q,\xi) = \sum_i c_i \left[\phi_0(q) + \sum_n \phi_n(q)\hat{g}_{ni}(\xi)\right]\psi_{0i}(\xi) \ , \tag{10}$$

$$\psi_{ni}(\xi) = \hat{g}_{ni}(\xi)\psi_{0i}(\xi) \equiv \int_{\Omega_\xi} d\xi' g_{ni}(\xi,\xi')\psi_{0i}(\xi') \ , \tag{11a}$$

$$g_{ni}(\xi,\xi') = \sum_{i'} \frac{\psi_{ni'}^0(\xi)V_{n0}(\xi')\psi_{ni'}^{0*}(\xi')}{\eta_i - \eta_{ni'}^0 - \varepsilon_{n0}} \ , \tag{11b}$$

where $\{\psi_{0i}(\xi)\}$ are the eigenfunctions and $\{\eta_i\}$ the eigenvalues found from the effective equation, eq. (7), while the coefficients $c_i$ should be determined from the boundary conditions understood as the state-function matching conditions along the boundary where $V_{eg}(q,\xi) = 0$. The observable, physical density, $\rho(q,\xi)$, is obtained then as the squared modulus of the ('classical') field amplitude: $\rho(q,\xi) = |\Psi(q,\xi)|^2$.

It is clear that eqs. (7)-(11) represent just a reformulation of the initial problem of interacting fields (any objects in the general case) which cannot directly provide a closed, 'exact' solution, since the effective interaction, eqs. (8), depends on the unknown solutions of the auxiliary system. This 'effective' formulation of a problem is well known in the theory of scattering, or in general as an approach to the many-body problem [27], under the name of 'optical potential method', where the optical potential is a synonym to our EP and is determined by expressions analogous to eqs. (8). However, those expressions are then always artificially simplified, using a version of perturbation theory, in order to obtain the solution in a closed form. The inevitable consequence is that such a *reduced* solution can only slightly deviate from the free-field (zero-interaction) solution, which is equivalent to the unrealistic (and uninteresting) weakness of the interaction effect. Because of the resonant structure of the denominators in the unreduced expression for EP, eq. (8b), it can have arbitrarily large magnitude for any, however small, coupling interaction. In reality, due to their self-consistent dependence on the eigenvalues to be found ($\eta$), the resonant denominators dynamically adjust themselves so as to satisfy the effective existence equation, eq. (7). It is precisely this qualitatively important EP dependence on the eigen-solutions of a problem that explicitly reveals the *universal* essentially *nonlinear* character of *any* interaction process hidden in the initial formulation, eqs. (1) or (4), and irreversibly cut off within canonical perturbative approaches looking for 'exact' solutions completely *separable* into one-dimensional components.[2] Now we can show that this particular phenomenon of the self-sustained, universal dynamic nonlinearity not only forms an insurmountable difficulty for the canonical approaches, but also provides an issue towards the truly complete solution within the unreduced effective interaction analysis revealing thus the genuine meaning of the *formally* equivalent problem transformation to the effective formulation, eqs.(7)-(11).

Namely, we show that the mentioned dynamic nonlinearity of the effective interaction leads to emergence of *multiple* solutions to a problem, each of them being 'complete' in the ordinary sense and therefore *incompatible* with other solutions which form thus a *redundant* set and, being equally valid, should permanently and 'spontaneously' *replace* each other in the observed system behaviour [23-25]. The *dynamic multivaluedness* of solutions directly follows from the effective formulation of a problem, eqs. (7)-(8), if we take into account the increased maximum power of the eigenvalue to be found, $\eta$, with which it enters the characteristic equation obtained from eq. (7) and incorporating the unreduced EP expression, eq. (8b). If $N_e$ and $N_g$ are the numbers of the eigenvalues for the free e/m and gravitational protofields respectively (usually $N_e = N_g$, and this number represents the number of the eventually emerging physical 'points of contact' of the interacting protofields, see also below), then the number of eigenvalues for the unreduced problem with interaction, $N_{eg}$, determined by the maximum power of $\eta$ in the characteristic equation for the effective generalised Hamiltonian of eq. (7), is

$$N_{eg} = N_g N^1_{eg} = N_g(N_e N_g + 1) , \qquad (12)$$

where the factor of $N^1_{eg} = N_e N_g + 1$ solutions comes from the dependence of the effective potential on $\eta$, eq. (8b) (the calculations are straightforward). Since the total number of eigen-solutions, $N^*_{eg}$, for the ordinary, non-modified formulation of the problem represented by the system of eqs. (4) is $N^*_{eg} = N_e N_g$, it becomes clear that the effective formulation reveals a hidden dynamic multiplication of the number of solutions forming an $N_g$-fold redundant set of ordinary *complete* sets of $N^*_{eg}$ solutions, each of them representing a 'normal', full version of the compound system called

---

[2]Therefore such essential, dynamically maintained nonlinearity, universally specified in the unreduced effective interaction analysis, is quite different from any formally introduced 'nonlinear dependence' that can be totally compatible with 'exact' solutions (cf. 'solitons', etc.).

*realisation* and differing from other redundant realisations in parameters determined by the corresponding version of the effective interaction. In addition to the $N_g$-fold redundance of those 'regular' realisations, the total number of solutions given by eq. (12) includes $N_g$ special eigen-solutions which form a specific 'intermediate', incomplete and therefore highly unstable realisation serving as a common, highly irregular 'transitional state' for the system performing spontaneous 'jumps' between the redundant 'regular' realisations [23-25]. The existence of the transitional realisation and irregular jumps is related to the fundamental difference of the revealed *dynamic* multivaluedness from the ordinary, unspecified 'nonuniqueness of solution' in the canonical approaches: although our dynamically redundant solution-realisations are mutually incompatible, they are *dynamically connected*, with the help of the unique intermediate realisation, in an *intrinsically continuous*, holistic sequence of natural stages forming a *single*, though internally highly *nonuniform*, dynamical process.[3]

An equivalent graphical analysis of the same phenomenon of dynamic redundance [23-25] only confirms the obtained results, as well as their universality and irreducible equivalence (reality) of the non-identical redundant solution-realisations, i. e. the fact that it is impossible to reduce 'excessive' solutions to some 'spurious' effects or purely mathematical 'artefacts'. Moreover, it becomes clear that it is precisely this irreducible and universal solution multivaluedness of purely dynamic origin that can uniquely account for the *causal* (dynamic) emergence of *intrinsic randomness* in the world, starting already from the most elementary levels of the coupled protofields dynamics, and provides the corresponding totally self-consistent, 'natural' definition of *probability*. If we choose index $j$ to enumerate the discovered 'complete' sets (realisations) of $N_{eg}^*$ eigenvalues and eigenfunctions of the modified (effective) existence equation, eqs. (7)-(8), then all other related quantities within the total solution, eqs. (10)-(11), also acquire the dependence on $j$ forming now a *really* complete *set of realisations*, $\mathfrak{R} = \{\mathfrak{R}_j\}$, each of them consisting of an ordinary 'complete' component, or 'version', of the variables of a problem: $\mathfrak{R}_j = (\{\eta_i\}^j, \{\psi_{0i}(\xi)\}^j, V_{\text{eff}}^j(\xi;\eta^j), \{\psi_{ni}(\xi)\}^j, \Psi_j(q,\xi), \rho_j(q,\xi))$, $j = 1,2,...,N_\mathfrak{R}$, where $N_\mathfrak{R}$ is the total number of realisations (as we have seen above, $N_\mathfrak{R} = N_g$). The dynamic redundance concept justified above states that the genuine (i. e. really complete) *general solution* of a problem, representing the above-mentioned unreduced complex-dynamical process, is obtained at the level of the *observable* (generalised) density $\rho(q,\xi)$ as a *causally probabilistic sum* of the densities, $\rho_j(q,\xi)$, for individual system realisations:

---

[3]Note that the ordinary 'uniqueness theorems' actually are not applicable to the unreduced dynamics, since they silently imply the singlevaluedness of the interaction potential, and this condition is 'self-consistently' violated for the multivalued effective dynamics.

$$\rho(q,\xi) = \sum_{j=1}^{N_{\Re}}{}^{\oplus} \rho_j(q,\xi) \ , \tag{13}$$

where the sign $\oplus$ serves to designate the special, *dynamically probabilistic* meaning of the sum. The latter implies that the individual, *completely defined* compound-system densities $\rho_j(q,\xi)$ *chaotically*, i. e. spontaneously/unpredictably, appear and disappear in the actual observations, *whatever* is the time of observation and the number of registered events (an *event* is *rigorously defined* as a dynamically spontaneous realisation emergence). Since the elementary realisation-events dynamically emerge in the system behaviour and its complete description on absolutely *equal* grounds, we obtain the *a priori determined* probability, $\alpha_j$, of the *j*-th elementary realisation emergence as

$$\alpha_j = \frac{1}{N_{\Re}} \quad (j = 1,...,N_{\Re}) \ , \quad \sum_j \alpha_j = 1 \ . \tag{14a}$$

Since in many practical cases the elementary realisations are not individually resolved in the actual measurements and, being inhomogeneously distributed, appear in dense groups containing various numbers of them, in the general case the obtained probabilities are not equal:

$$\alpha_j(N_j) = \frac{N_j}{N_{\Re}} \quad (N_j = 1,...,N_{\Re}; \ \sum_j N_j = N_{\Re}) \ , \quad \sum_j \alpha_j = 1 \ , \tag{14b}$$

$N_j$ being the number of elementary realisations in their *j*-th group. When the number of events (observation time) is large enough, one measures the familiar *expectation (average) value*,

$$\rho_{\text{ex}}(q,\xi) = \sum_{j=1}^{N_{\Re}} \alpha_j \rho_j(q,\xi) \ . \tag{15}$$

However, the dynamically probabilistic sum of eq. (13) and the associated causally deduced probability values, eqs. (14), preserve their meaning also for *single, isolated* events emerging in the *real time* of observation/measurement (and thus *together* with the time itself, see below).

Note that the observed probabilistic sum of densities of *incompatible* individual realisations is the causal and *universal* extension of the canonical notion of the 'density matrix' [25] which otherwise would necessarily imply the existence of a basically 'unknown' influence of some additional 'environment' [1] and cannot clearly specify the *physical* difference between the 'pure' and 'mixed' states. Correspondingly, the incompatibility of the *physically specified* realisations is quite different from (often ambiguous) 'incoherence', or 'decoherence', of the abstract *superposable* states in the canonical 'interpretations'.[4] The dynamic redundance should also be clearly distinguished from purely speculative assumptions of the well-known 'many worlds interpretations', since in our approach we insist on existence of *only one* tangible reality, which just explains the necessity of permanent chaotic *change* of realisation currently occupying that *unique* place, while all the other *locally* determined realisations, or 'versions of reality', are 'trying to enter' into that unique place from their (transient) 'potential' state within the above intermediate (transitional) realisation of a system with interaction.

The *dynamic complexity* of a system as such can be defined as a simple increasing, positively defined function of the number of realisations, equal to zero for the exceptional case of only one realisation of a *regular* system, or equivalently, as a rate of realisation change by the system [25] (see also further development of our analysis). The rigorously defined *chaos* is considered here as a synonym of the unreduced dynamic complexity (redundance), and *chaotic* means dynamically complex (multivalued). Chaos thus defined is always 'dynamical', which means, specifically, that *every* elementary jump between realisations is causally unpredictable by its exact result, but the *distribution* of realisation *probabilities* is completely defined and regular (and can be calculated). Therefore chaos is always a *combination* of the *regular*, global (average) tendency, represented by the probability distribution, and *purely irregular* deviations from that tendency, representing the causal randomness as such. Correspondingly, a chaotic behaviour is always a *partially* ordered randomness ('order in chaos'), or partially disordered (probabilistically emerging) regularity.

Now in order to understand the nature of the resulting physical *configurations* of the individual realisations and their probabilistic mixture in the general solution of eq. (13), we use eqs. (10)-(11) to write down explicitly the expression for the measured density of the *j*-th 'pure' state-realisation entering that mixture

---

[4]The purely *operational* canonical definition of *coherence* between components of a 'pure' state and *incoherence* (decoherence) between those of a 'mixed' state is formulated in terms of rules of *superposition* of the corresponding purely abstract 'Hilbert space vectors' of the component states that should be summed up as vectors before finding the measurable squared modulus, for the coherent/pure state, or else should be summed up by their respective squared modulus, in the case of decoherent/mixed state. The dynamic redundance concept replaces the ambiguous technical manipulations by *dynamically emerging* probabilistic 'switching', eq. (13), between *incompatible* states/realisations of *physically real* (coupled) fields (instead of *superimposed* 'incoherent' vector components), while each realisation is an internally 'coherent' state of the same physical fields.

$$\rho_j(q,\xi) = |\Psi_j(q,\xi)|^2, \quad \Psi_j(q,\xi) = \qquad (16)$$

$$= \sum_i c_i^j \left[ \phi_0(q)\psi_{0i}^j(\xi) + \sum_{n,i'} \frac{\phi_n(q)\psi_{ni'}^0(\xi) \int_{\Omega_\xi} d\xi' \psi_{ni'}^{0*}(\xi')V_{n0}(\xi')\psi_{0i}^j(\xi')}{\eta_i^j - \eta_{ni'}^0 - \varepsilon_{n0}} \right].$$

It is easy to see from this expression that the compound field magnitude for each realisation, represented by $\Psi_j(q,\xi)$, is *physically* concentrated around certain 'preferred' eigenvalue, while gradually decreasing with distance from it, measured in the space formed by eigenvalues [24,25].[5] The preferred eigenvalue for the *j*-th realisation is given approximately by members of the found set $\{\eta_i^j\}$ for which the denominator and numerator have generally close values which are further adjusted through similar eigenvalue dependence of EP, eqs. (8). It is important that the latter has the same form as the above state-function expression:

$$V_{\text{eff}}(\xi;\eta_i^j)\psi_{0i}^j(\xi) = V_{00}(\xi)\psi_{0i}^j(\xi) + \sum_{n,i'} \frac{V_{0n}(\xi)\psi_{ni'}^0(\xi) \int_{\Omega_\xi} d\xi' \psi_{ni'}^{0*}(\xi')V_{n0}(\xi')\psi_{0i}^j(\xi')}{\eta_i^j - \eta_{ni'}^0 - \varepsilon_{n0}}.$$

(8c)

This means that the effective interaction for the *j*-th realisation has the largest amplitude, i. e. dynamically produces the potential well, just around the same eigenvalue for which the state-function for that realisation has its 'centre of concentration', which clarifies the physical meaning and demonstrates the reality of the 'effective' description of a problem, where a 'potential well' is *dynamically formed* within the *same* nonlinear reduction process as the state 'trapped' in it.

---

[5]Although we shall not analyse here the detailed law of that concentration of the state-function within a regular realisation, it is evident that the denominators in eq. (16) will give the inversely proportional dependence of $\Psi$ on the 'generalised (realisation) coordinate', while the 'matrix elements' in the numerator, determined by superposition of the participating localised functions, should introduce a much stronger, exponential dependence. In the case of the dynamically squeezed elementary field-particle (see below), this kind of dependence corresponds to the one obtained by de Broglie within the original double-solution theory [12,13], where he used, however, an externally inserted, model nonlinearity instead of our self-consistently (dynamically) appearing, universal 'feedback' nonlinearity of EP dependence on the eigen-solutions.

If we apply this universal interpretation more specifically to the considered case of interacting e/m and gravitational protofields where the eigenvalues, such as $\eta_i^j$, $\eta_{ni'}^0$, and $\varepsilon_{n0}$, are chosen to be the (emerging) space 'coordinates' (see the remark after eqs. (2)-(3)), then we see that the unreduced dynamics of the system can be described as unceasingly repeated stage of a nonlinear *dynamical squeeze* of the e/m field to unstable localised, *corpuscular* state-realisations of thus emerging (physical) 'points of space' (or 'centres of reduction') represented by their central ('preferred') *coordinates* $\eta_i^j$ (they are practically degenerate with respect to the local 'excited states' enumerated by index *i*) which alternates with the opposite stage of the field *extension* to a transient delocalised, *wave-like* state (the latter represents precisely the special 'intermediate' realisation described above that forms the causally extended, physically real *wavefunction* of the system, see also quant-ph/9902016). The gravitational degrees of freedom, being dynamically *entangled* with the e/m protofield, eq. (16), also participate in this never ending series of cycles of *causal (dynamical) reduction*, or *collapse*, and extension of the state-function $\Psi(q,\xi)$, but as experience shows, they remain *directly* unperceivable, forming a kind of 'dynamical (material) matrix' of the externally purely 'electromagnetic' world (while the *indirect* manifestation of the gravitational protofield is known as universal gravitation, see quant-ph/9902016, gr-qc/9906077). That gravitational medium-matrix performs thus the *generalised dynamical measurement* of the e/m protofield taking the form of the above universal reduction/extension process.

Each 'regular' realisation corresponds to a particular emerging centre of reduction, and since the realisations are redundant (incompatible), the system performs each reduction to a *causally random* point within the accessible (generally large) domain (it is also limited by a finite speed of propagation of the corresponding e/m protofield perturbations). In this way the system of two *homogeneously* interacting protofields dynamically *produces*, or 'weaves' the tissue (structure) of the *physically perceivable, material* entity of *space* in the form of *randomly* occupied points/centres of the highly *inhomogeneous* e/m field concentration in the corpuscular phase of collapse. Each such *physical space point* has a very small, but *finite dimension* determined by the balance, attained at the maximum squeeze moment, between the attractive coupling force and the finite compressibility of the protofields (this elementary spatial scale of the world seems to be of the order of a *renormalised* Planckian scale, around $10^{-17}$ cm, appearing as the smallest internal structure size of the elementary particle [25], gr-qc/9906077).

Similarly, the elementary, indivisible unit of *time* causally *emerges* as the period of one cycle of reduction and extension, whereas the *spatially* chaotic sequence of *events* of reduction/extension forms the continuous (though *internally* 'jumping')

*flow of time* (that is why time does not enter the realisation definition, eq. (16); it makes an elementary leap when one such realisation is replaced by another). However, contrary to space, the causal time though being (physically) *real* is *not* a material, tangible entity, or 'dimension': an elementary leap of time *shows* that an elementary *event* (of reduction/extension) has happened, it describes a *mode* of existence of an entity, and not its tangible *properties* (or *quality*) appearing as its *spatial structure* (or 'texture'). The period of the elementary cycle of reduction/extension should be *constant* and very small as it forms the finest, indivisible 'rhythm' (within the most accurate 'clock') of the world; it is determined by the magnitude of the electro-gravitational coupling and will be specified in further description (see quant-ph/9902016). Since we deal with the most fundamental level of the *closed* world dynamics, there should be no any 'dissipation', and the periodic sequence of reduction/extension cycles can never stop, which provides a transparent causal explanation for the ever *persisting*, unceasing character of the time *flow* (if the world is not absolutely isolated, the velocity of the time flow could slowly, 'adiabatically' vary). Another indispensable property of the flow of time, its *irreversibility*, is the direct *consequence* of the *causally random* choice of each next realisation, in the form of a centre of reduction (or dynamically woven space point): the intrinsic randomness *excludes*, and even makes *senseless*, any possibility of a guaranteed, exact 'backward' reproduction of any part of the process, starting from a single cycle. Therefore the entity of *time* is inseparable from the causal, dynamically based *randomness* (*probabilistic* character) of system dynamics: both properties emerge together with the unreduced, complex dynamics of the interaction process.

It is easy to see also that the obtained *dynamically discrete*, or *quantized* character of space and time, emerging in the initially *uniform* system with interaction, is due to the *globality* (wholeness) of complex behaviour within the intrinsically *complete* description of the dynamic redundance paradigm: the *unreduced* dynamics of a system where 'everything interacts with everything' *cannot* be subdivided *in principle* into parts or stages smaller than our causally derived realisations [25]. Referring to this dynamical quantization of the elementary field dynamics we shall call the whole described complex-dynamical process of electro-gravitational interaction *quantum beat*. It can also be presented as *spatially chaotic* (dynamically irregular) *wandering* of the dynamically squeezed, *corpuscular* field state which therefore can also be called *virtual* (= highly unstable) *soliton* and represents the causally derived version of de Broglie's 'hump' performing its wandering by periodic *transformation* into the extended, *wave-like* state of the intermediate realisation. This Brownian type of motion of the 'particle within its wave' is qualitatively very close to de Broglie's notion of the 'hidden thermodynamics of the isolated particle' [14-17], but now it does not need any external 'motor' of chaoticity, or 'hidden thermostat' (represented by a 'subquantum medium'). The intrinsically *dynamical* chaos *within* the quantum beat process provides also a natural combination of the described causal randomness with a partial *order* in the spatial distribution of the reduction *probability* accounting for the *global displacement* of the thus emerging *elementary field-particle* [25] (see also the next paper, quant-ph/9902016) which has the properties of the experimentally observed (truly) 'elementary particle'. Therefore, we can say that eqs. (13)-(16) together with the eigen-solutions of eqs. (7)-(8) constitute the basis of the explicit presentation of the causally complete *double solution with chaos* directly extending and unifying de Broglian ideas about the nonlinear wave mechanics and hidden thermodynamics of the elementary particle (we continue to specify it in further analysis, quant-ph/9902016; see also gr-qc/9906077, quant-ph/9911107). Due to the discovered dynamical redundance phenomenon, the obtained solution provides a mathematically consistent and physically transparent explanation of the 'mysterious' *wave-particle duality* which emerges as a 'normal', inevitable property of the unreduced *complex dynamics* of the two interacting protofields. In particular, the *permanent*, and always somewhat ambiguous, *superposition* of the two dual parts in the original version of the double solution is now replaced by their *unceasing* dynamical (and causally random) *alternation* (quantum beat), which immediately removes all the contradictions. The quantum-beat dynamics has a universally *nonunitary* character meaning that, contrary to any dynamical evolution within the single-valued paradigm of the canonical science, it is *qualitatively* nonuniform (dynamically quantized into incompatible state-realisations) and causally probabilistic (and therefore irreversible), both properties resulting directly and uniquely from the dynamic multitude of locally 'complete' realisations.

The discovered purely dynamic, first-principles origin of randomness in an elementary, closed, deterministic, and a priori homogeneous system with interaction permits us to avoid the evident contradictions and inevitable non-universality of various existing 'dynamical collapse' (or 'continuous measurement') theories invariably developed within the same single-valued, perturbational approach and therefore obliged to evoke a noisy 'influence of environment' on the *abstract* 'state vector' as the 'ultimate' source of randomness and reduction (e. g. [8, 28-30]; further details and relevant references can be found in [25]). The logical inconsistency ('vicious circle') of the 'density matrix' ('distribution function') type of formalism, often used by those unitary imitations of complexity (instead of a more fundamental, *wavefunctional* description) and *already* (tacitly) implying the result to be deduced, in combination with the practically arbitrary play with multiple parameters of an ambiguous, infinitely conforming 'influence of environment', permits one to 'obtain' any desired result and thus 'explain' everything. The concept of naturally *emerging* dynamic redundancy, consistently *derived* from the first principles, provides instead a qualitatively new, causally complete (and technically correct) extension of quantum mechanics by resolving the fundamental contradiction between the irreducibly probabilistic, internally changeable character of quantum phenomena and the formally imposed single-valuedness of the conventional unitary approaches (recall that single-valuedness of the wavefunction, and thus of the whole formalism, constitutes one of the canonical 'postulates' of the conventional quantum mechanics, actually accepted by all its existing 'interpretations', including the latest pretensions for the 'ultimate understanding' with the help of *externally* diverse mechanistic trickery around 'decoherence', 'quantum logic', 'spontaneous collapse', 'consistent histories', 'Bohmian mechanics', etc.).

Several important refinements should be added to the described results of the unreduced approach. We have already mentioned above that the elementary reduction-squeeze of the interacting protofields, forming the basis for the fundamental quantum-beat dynamics, has a self-sustained character. The latter results from the *universal dynamic nonlinearity*, in the form of *feedback loops* formed by the unreduced interaction itself and adequately described by the 'self-adjusted' EP dependence on the final eigen-solutions, eqs. (8). This is a physically transparent effect: the more is the concentration of the interacting fields around a randomly appeared fluctuation of increased density, the more is their local attraction to each other, so that even an 'infinitesimal' fluctuation will grow. We obtain thus the general mechanism of the basic *dynamic instability* of *any* real system with interaction (with respect to splitting into realisations) and of any one of the emerging state-realisations, including the extended 'transitional' state of the fields (the causally extended 'wavefunction', see quant-ph/9902016) which *should* be unstable and irregularly changing simply because of the plurality of redundant realisations. This causally specified, unreduced nonlinearity/instability of any interaction process, including the quantum beat dynamics, provides the natural completion for de Broglie's ideas about essential role of irreducible nonlinear effects in the detailed picture of quantum (wave) dynamics [12,13,17,18].

The self-amplified, *catastrophic* reduction of the whole system with interaction to particular randomly chosen realisations (like physical space points) is inseparable from the phenomenon of *dynamic entanglement* of the interacting entities. Indeed, the above physical considerations and corresponding mathematical expressions refer to *all* interaction partners (*both* interacting fields, in our case). It becomes clear then that they should inevitably gradually (physically) *interlace* with each other, in course of the catastrophic squeeze-reduction and within every *single* realisation of the thus emerging *physically compound* system. The 'free', non-interacting protofields contain, respectively, $N_e$ and $N_g$ operational eigen-modes (prototypes of the emerging physical 'space points', where normally $N_e = N_g$). Their *unreduced* interaction includes $N_e N_g$ 'entangled' combinations, whereas there is always the same number of only $N_e$ 'places' for the 'entangled' modes which exist in the *same reality* as the free (e/m) field. This provides a transparent demonstration of the origin and mechanism of dynamic redundance of the formed $N_g$ incompatible states-realisations, each of them containing an inseparable mixture of $N_e$ 'free-field' modes of both protofields. The resulting physical, emergent *mixing*, or entanglement, is thus another necessary aspect of dynamic complexity, closely related to realisation redundance and described by products of elementary states for each of the interacting entities (depending on $q$ and $\xi$ respectively) in the expressions for the general solution, eqs. (10), (16). Correspondingly, the reverse process of extension of the interacting protofields towards the 'transitional', delocalised realisation is accompanied by their temporal, unstable *disentanglement*, so that the fields transiently return to their 'quasi-free', uncoupled state, which is inevitable because of the *omnipresent* character of the driving protofield attraction and physically necessary in order that they can form the next realisation with *another* detailed configuration of entanglement in course of the next *causally random* reduction. This dynamically maintained alternation of intrinsically probabilistic events of entanglement and disentanglement reveals the universal internal structure of *any* interaction process [25] always giving the sequence of changing phases of 'strong' and 'weak' *effective* interaction, with respect to the 'average' magnitude of initial coupling.

Moreover, it is extremely important that the dynamic entanglement (and disentanglement) does not stop there and continues to develop in the same fashion and by the same mechanism to ever finer scales, forming eventually what we call the *fundamental dynamical fractal* of a problem (system). Indeed, we have restricted our explicit analysis to dynamical splitting of the solution of the effective existence equation, eq. (7), while using the solutions of the auxiliary system of equations (9) in their general, unspecified form. However, in reality those solutions, $\{\psi_{ni}^0(\xi)\}$, $\{\eta_{ni}^0\}$, being a part of the solution for the whole problem, depend themselves on the eigenvalues to be found. We can explicitly take into account this dependence by applying the *same* EP method now to the auxiliary system which naturally gives further dynamical splitting of each of the previously obtained realisations into redundant realisations of the second level expressed through solutions of another truncated system of equations [25]. Repeating the same procedure, we obtain a practically unlimited hierarchy of levels of dynamically redundant realisations of ever finer scale which forms the mentioned dynamical fractal of a problem. The branches-realisations at any level of this universal dynamical arborescence are obtained as a result of the physically real entanglement of the interacting entities, quite similar to the one of the first level presented above. Because of the intrinsic incompatibility of realisations, the system never stops changing them, at any level, and therefore this arborescent fractal structure of complex dynamics preserves its *probabilistic*, ever changing, or 'living' (self-developing, dynamically adaptable) character. The high involvement of the resulting physical structure demonstrates the unreduced intricacy of complex dynamics and represents actually the general solution of a problem in its explicit form (the general solution of eqs. (10), (16) shows explicitly only the first level of the hierarchy, while implicitly containing all of them). We shall not present here the detailed expressions of this analysis of dynamic fractality which is a rather straightforward version of the above results [25]; besides, they are indeed less important than the qualitative picture substantiated above, since the *quantized* reduction-extension events can be observed only *as a whole* at this most fundamental level of the world dynamics. One should just keep in mind that each elementary reduction-extension event within the quantum beat process, including its corpuscular stage (i. e. the 'elementary particle' as such), has its involved internal structure containing a dynamically branching, fractal hierarchy of randomly formed, naturally entangled configurations of the e/m and gravitational protofields which are exhaustively described by the unreduced effective formalism, eqs.(7)-(8), (13)-(16) (and its straightforward extension to finer scales).[6]

The described picture of dynamic entanglement extends also the formal notion of *nonseparability* of the essentially linear (unitary) science: *any* generic system with interaction is nonseparable, *both physically and mathematically*, because the interacting physical entities *really* (and autonomously) entangle with each other in a (practically) *infinitely* fine network. The close notion of *nonintegrability* can be considered as a direct consequence of the dynamic multivaluedness: *any* generic system is 'nonintegrable' by the methods of unitary science because it always has *many* equally probable solutions, whereas it is supposed to have *only one* (exhaustive) solution by the unitary paradigm of single-valuedness. The opposite notions of separability and integrability are straightforward (unrealistic) exceptions from those generic cases, which provides also a universal practical test for the corresponding properties proceeding through definition of the operational number of system realisations by the above universal analysis (an integrable system has only one effective realisation and is characterised by an effectively *one-dimensional*, single-valued, regular behaviour).

---

[6]Note that our dynamical fractal is explicitly obtained as a complete, complex-dynamical *solution* for a general problem of a system with interaction, possessing an *intrinsically* probabilistic character, spatial discreteness, and temporal irreversibility, whereas the canonical, single-valued science proposes only mechanistic imitations of fractality which are obtained either by semi-empirical simulations, or by artificial insertion of the necessary properties estimated then by the resulting formal 'signatures'. Similarly, the notion of 'quantum entanglement' of the *unitary* quantum mechanics, very actively applied now especially in discussion of 'quantum information processing', is completely devoid of any realistic, dynamical basis and the ensuing intrinsic properties of wholeness, irregularity, and irreversibility. This major deficiency leads, in particular, to the completely erroneous conclusion about the feasibility of the unitary 'quantum computation' possessing a 'miraculously' increased power (see [25] for more detail).

We cannot avoid mentioning here another closely related feature of the complex quantum-beat dynamics giving rise to the causal understanding of the universal property of *spin* of an elementary field-particle (see [25] for a more detailed analysis). The picture of the dynamical, physically real interlacement of the interacting protofields already implies a clear image of the necessary rotational component of motion of the fields around each reduction center. This tentative image can be provided with a fundamental substantiation if we notice that the self-sustained squeeze of an *extended* protofield with a *finite* compressibility to a *very small* volume should inevitably result in the shear instability within the squeezing field leading to its *vortex-like motion* towards the centre of reduction. The phenomenon is generally similar to the (turbulent) forced passage of a liquid through a small hole, but in our case we have its much more 'nonlinear', self-amplified version, while the *unique* role of the *complex* quantum-beat dynamics consists in 'automatic', *unceasing* creation of a *sequence* of those 'holes' (any canonical, unitary theory would predict for this case a single, more or less homogeneous 'fall' of one of the protofields onto another). It is this highly nonuniform vortex motion of the elementary field-particle in course of *each* reduction-extension cycle which provides both causal, universal understanding of the property of spin, and further refinement of the process of dynamical entanglement of the interacting (physical) degrees of freedom within the holistic quantum beat process.

Finally, we can specify the boundary conditions mentioned above (after eqs. (10), (11)) to show that the obtained values of coefficients $c_i^j$ in the general solution (see also eq. (16)) are equivalent to the rules for probabilities of eqs. (14) deduced above from general logic. Note that analysing the unreduced complex behaviour, we always deal with a new type of 'dynamical boundary (or initial) conditions'. It follows already from the property of intrinsic continuity of complex dynamics presented before which does not permit one to fix some artificially chosen, 'geometrical' boundary manifold situated somewhere 'before the interaction has started', as it is usually done in the single-valued, unitary science. In return, within the universal complex-dynamical process of realisation change we have another kind of 'surface' where the interaction effectively vanishes: it is the 'intermediate', or 'transitional' realisation in which the interacting entities transiently disentangle. This realisation can also be considered as the 'initial moment of time' in the 'temporal formulation' of a problem, where the entity of 'time' actually just emerges within the same process of realisation change. Matching now the system state-function for the intermediate realisation, $\Psi_0(q)$, with the one for certain $j$-th localised realisation-point, eq. (16), and averaging the internal intensity distribution by integration within the closed vicinity of the reduction centre, we get

$$\Psi_0(q=q_j) \equiv \Psi_0 = \sum_i c_i^j \equiv C_j , \qquad (17a)$$

where we have taken into account that the intermediate realisation is homogeneous in our case and used renormalised (averaged) values of the coefficients $c_i^j$. It is easy to see now that the latter can be presented as products, $c_i^j = C_j c_i$, where the coefficients $c_i$ are responsible for less important details of the internal distribution of density of the localised (corpuscular) state of the field-particle, while the coefficients $C_j$ should be identified with the 'amplitudes' of the probabilities $\alpha_j$ introduced above, so that

$$\alpha_j = |C_j|^2 = \frac{1}{N_\Re} \;\; (j = 1,...,N_\Re), \;\; \sum_j \alpha_j = 1 , \;\; \sum_i c_i = 1 . \qquad (17b)$$

Note that the found correspondence, eqs. (17), between the state-function values of the quasi-free 'measured' object (e/m protofield) in the intermediate, extended realisation and the respective probabilities of individual (localised) realisation emergence remains valid, together with the whole above analysis, for any system with interaction. In particular, if the 'measured' interaction partner has an arbitrary inhomogeneous distribution of its magnitude, then the distribution of realisation probabilities should reproduce exactly this inhomogeneous configuration (see eq. (17a)). This law, causally *derived* here within the unreduced description of complex dynamics, can be called 'generalised Born's probability rule', since at the next higher level of the observed quantum field mechanics, i. e. that of the Schrödinger equation (see the second paper, quant-ph/9902016), it gives the same result as the *postulated* introduction of probability into the standard quantum mechanics proposed by Born. We emphasize the qualitative, and physically transparent, basis of our causal extension of the probability rule within the 'complex-dynamical boundary/initial conditions': since the incompatible realisations are *directly*, physically 'produced', in course of the intrinsically *continuous* dynamical reduction-squeeze, from the 'free' interacting entities in the intermediate realisation, transiently reproducing the 'initial' state, the probability distribution of their causally random emergence *should* coincide with the system density (intensity) distribution in that intermediate (or initial) state-realisation.

Now that the basic properties of the complex dynamics of interacting protofields are specified in the described form of the double solution with chaos (quantum beat), one can turn to the main observed consequences of this complex behaviour. They appear at higher levels of complexity progressively emerging as a result of elementary-realisation interaction through their fractal network at smaller scales, always within the same universal mechanism of dynamical splitting into new redundant realisations (in this way the fundamental dynamical fractal continuously develops itself both 'in deep', to finer scales, and to higher levels of complexity). It will be shown in the next paper (quant-ph/9902016) that this natural complexity development gives the causally substantiated, internally chaotic Schrödinger (Dirac) dynamics of the intrinsically quantized wave field of the quantum beat process, which is naturally unified with the causally extended relativity of the corpuscular-state motion and passes to classical type of behaviour for systems with higher complexity (starting from elementary bound systems, like atoms).